
\input harvmac
\noblackbox
\newcount\figno
\figno=0
\def\fig#1#2#3{
\par\begingroup\parindent=0pt\leftskip=1cm\rightskip=1cm\parindent=0pt
\baselineskip=11pt
\global\advance\figno by 1
\midinsert
\epsfxsize=#3
\centerline{\epsfbox{#2}}
\vskip -3cm
\vskip 12pt
\endinsert\endgroup\par}
\def\figlabel#1{\xdef#1{\the\figno}}
\def\pano{\par\noindent}
\def\smno{\smallskip\noindent}
\def\meno{\medskip\noindent}

\font\cmss=cmss10
\font\cmsss=cmss10 at 7pt
\def\rlx{\relax\leavevmode}
\def\inbar{\vrule height1.5ex width.4pt depth0pt}
\def\IC{\relax\,\hbox{$\inbar\kern-.3em{\rm C}$}}
\def\IR{\relax{\rm I\kern-.18em R}}
\def\IN{\relax{\rm I\kern-.18em N}}
\def\IP{\relax{\rm I\kern-.18em P}}
\def\ZZ{\rlx\leavevmode\ifmmode\mathchoice{\hbox{\cmss Z\kern-.4em Z}}
 {\hbox{\cmss Z\kern-.4em Z}}{\lower.9pt\hbox{\cmsss Z\kern-.36em Z}}
 {\lower1.2pt\hbox{\cmsss Z\kern-.36em Z}}\else{\cmss Z\kern-.4em Z}\fi}
\def\narrowplus{\kern -.04truein + \kern -.03truein}
\def\narrowminus{- \kern -.04truein}
\def\narrowminussub{\kern -.02truein - \kern -.01truein}

\def\cl{\centerline}

\def\o#1{\overline{#1}}

\def\th#1#2{\vartheta\bigl[{  #1 \atop #2} \bigr] }

\def\ra{\rangle}

\def\rN{\rm N}

\def\frac#1#2{{#1\over #2}}
\def\sqr#1#2{{\vcenter{\vbox{\hrule height.#2pt
 \hbox{\vrule width.#2pt height#1pt \kern#1pt
 \vrule width.#2pt}\hrule height.#2pt}}}}


\def\drawbox#1#2{\hrule height#2pt 
        \hbox{\vrule width#2pt height#1pt \kern#1pt 
              \vrule width#2pt}
              \hrule height#2pt}

\def\Asym#1#2{\vcenter{\vbox{\drawbox{#1}{#2}
              \kern-#2pt       
              \drawbox{#1}{#2}}}}


\lref\rdienes{J.D. Blum and  K.R. Dienes, {\it Duality without Supersymmetry: 
The Case of the SO(16)$\times$SO(16) String}, 
Phys.Lett. {\bf B414} (1997) 260, hep-th/9707148;
{\it Strong/Weak Coupling Duality Relations for Non-Supersymmetric String 
Theories}, Nucl.Phys. {\bf B516} (1998) 83, hep-th/9707160.}

\lref\ads{I. Antoniadis, E. Dudas and A. Sagnotti,
{\it Supersymmetry Breaking, Open Strings and M-theory},
Nucl.\ Phys.\ {\bf B544} (1999) 469, hep-th/9807011.}

\lref\aads{I. Antoniadis, G. D'Appollonio, E. Dudas and A. Sagnotti,
{\it Partial Breaking of Supersymmetry, Open Strings and M-theory},
Nucl.\ Phys.\ {\bf B553} (1999) 133, hep-th/9812118.}

\lref\adsa{I. Antoniadis, E. Dudas and A. Sagnotti,
{\it Brane Supersymmetry Breaking}, hep-th/9908023.}

\lref\aq{I. Antoniadis and M. Quiros,
{\it Supersymmetry Breaking in M-theory
and Gaugino Condensation},
Nucl.\ Phys.\ {\bf B505} (1997) 109, hep-th/9705037;
{\it On the M-theory Description of Gaugino Condensation},
Phys.\ Lett.\ {\bf B416} (1998) 327, hep-th/9707208.}

\lref\horava{P. Horava,
{\it Gluino Condensation in Strongly Coupled Heterotic String Theory},
Phys.\ Rev.\ {\bf D54} (1996) 7561, hep-th/9608019.}

\lref\rkks{S. Kachru, J. Kumar, E. Silverstein, {\it Vacuum Energy
Cancellation in a Non-supersymmetric String}, Phys. Rev. {\bf D59} (1999)
106004, hep-th/9807076; S. Kachru and E. Silverstein, {\it Self-Dual 
Nonsupersymmetric Type II String Compactifications}, JHEP 
{\bf 9811} (1998) 001, hep-th/9808056;
S. Kachru and E. Silverstein,
{\it On Vanishing Two Loop Cosmological Constant in Nonsupersymmetric 
Strings},  JHEP {\bf 9901} (1999) 004, hep-th/9810129.}

\lref\rshiutye{G. Shiu and S.-H.H Tye, {\it Bose-Fermi Degeneracy and 
Duality in Non-Supersymmetric Strings}, Nucl.Phys. {\bf B542} (1999) 45,
hep-th/9808095.}

\lref\rharv{J. A. Harvey, {\it String Duality and Non-supersymmetric Strings},
Phys.Rev. {\bf D59} (1999) 026002,  hep-th/9807213.}

\lref\rblgo{R. Blumenhagen, L. G\"orlich, 
{\it Orientifolds of Non-Supersymmetric 
Asymmetric Orbifolds}, Nucl.Phys. {\bf B551} (1999) 601, hep-th/9812158.}

\lref\raaf{C. Angelantonj, I. Antoniadis, K. F\"orger, 
{\it Non-Supersymmetric Type I 
Strings with Zero Vacuum Energy}, hep-th/9904092.}

\lref\sat{B. Sathiapalan, 
{\it Vortices on the String World Sheet and Constraints on Toroidal
Compactification},
Phys.\ Rev.\ {\bf D35} (1987) 3277.}

\lref\kog{I. Kogan,
{\it Vortices on the World Sheet and String's Critical Dynamics},
JETP Lett.\ {\bf 45} (1987) 709.} 

\lref\wittena{E. Witten,
{\it Anti-de Sitter Space and Holography}, Adv.\ Theor.\ Math.\ Phys.\
{\bf 2} (1998) 253, hep-th/9802150.}

\lref\wittenb{E. Witten,
{\it Anti-de Sitter Space, Thermal Phase Transition and Confinement
in Gauge Theories},
Adv,\ Theor,\ Math.\ Phys.\ {\bf 505} (1998) 505, hep-th/9803131.}

\lref\wittenc{E. Witten,
{\it String Theory Dynamics in Various Dimensions},
Nucl.\ Phys.\ {\bf B443} (1995) 85, hep-th/9503124.}

\lref\gkp{S.S. Gubser, I.R. Klebanov and A.M. Polyakov,
{\it Gauge Theory Correlators from Noncritical String Theory},
Phys.\ Lett.\ {\bf B428} (1998) 105, hep-th/9802109.}

\lref\adln{M. Abou-Zeid, B. de Wit, D. L{\"u}st and H. Nicolai,
{\it Space-time Supersymmetry, IIA/B Duality and M-Theory},
hep-th/9908169.}

\lref\dhs{M. Dine, P. Huet and N. Seiberg,
{\it Large and Small Radius in String Theory},
Nucl.\ Phys.\ {\bf B322} (1989) 301.}

\lref\dlp{J. Dai, R.G. Leigh and J. Polchinski,
{\it New Connections between String Theories},
Mod.\ Phys.\ Lett.\ {\bf A4} (1989) 2073.}

\lref\scherk{J. Scherk and J.H. Schwarz,
{\it Spontaneous Breaking of Supersymmetry through Dimensional Reduction},
Phys.\ Lett.\ {\bf 82B} (1979) 60.}

\lref\rKP{ C. Kounnas and M.  Porrati,
{\it Spontaneous Supersymmetry Breaking in String Theory}
 Nucl.\ Phys.\ {\bf B310} (198) 355.}

\lref\fkpz{S. Ferrara, C. Kounnas, M.  Porrati and F. Zwirner,
{\it Superstrings with Spontaneously Broken Supersymmetry 
and their Effective Theories}, Nucl.\ Phys.\ {\bf B318} (1989) 76.}

\lref\rschw{ J.H. Schwarz, {\it The Power of M Theory}, 
Phys. Lett. {\bf B367} (1996) 97,
hep-th/9510086; J.H. Schwarz, {\it Lectures on Superstring and M Theory 
Dualities}, Nucl. Phys. Proc. Suppl. {\bf B55} (1997) 1, hep-th/9607201.}

\lref\koubps{C. Kounnas, {\it BPS States in Superstrings with Spontaneously
Broken SUSY},
Nucl.\ Phys. Proc.Suppl.{\bf B58} (1997) 57.} 

\lref\KK {E. Kiritsis and  C. Kounnas, {\it Perturbative and Nonperturbative
Supersymmetry Breaking:  N=4 $\to$ N=2 $\to$ N=1}
Nucl.\ Phys.\ {\bf B503} (1997) 117.}

\lref\kns{I.R. Klebanov, N.A. Nekrasov and S.L. Shatashvili,
{\it An Orbifold of Type 0B Strings and Non-supersymmetric Gauge Theories},
hep-th/9909109.}

\lref\ratwitt{J.J. Attick and E. Witten,
{\it 
The Hagedorn Transition and the Number of Degrees of Freedom of String Theory},
Nucl.\ Phys.\ {\bf B310} (1988) 291.}

\lref\rkonrost{C. Kounnas and B. Rostand,
{\it Coordinate Dependent Compactification and Discrete Symmetries},
Nucl.\ Phys.\ {\bf B341} (1990) 641.}

\lref\rAK{I. Antoniadis and  C. Kounnas,
{\it Supersymmetric Phase Transition at High Temperature},
Phys.\ Lett.\ {\bf B261} (1991) 369.}
 
\lref\ADKa{I. Antoniadis, J.P. Derendinger and C. Kounnas,
{\it Nonperturbative Temperature Instabilities in N=4 Strings}
Nucl.\ Phys.\ {\bf B551} (1999) 41. }

\lref\ADKb{I. Antoniadis, J.P. Derendinger and C. Kounnas,
{\it Nonperturbative Supersymmetry Breaking and Finite Temperature
Instabilities in N=4 Superstrings}, hep-th/9908137.}

\lref\rbfla{R. Blumenhagen, A. Font and D. L\"ust,
{\it Non-supersymmetric Gauge Theories from D-branes in Type 0 String Theory},
hep-th/9906101.}

\lref\rluthzoup{D. L\"ust, S. Theisen and G. Zoupanos,
{\it Four-dimensional Heterotic Strings and Conformal Field Theory},
Nucl.\ Phys.\ {\bf B296} (1988) 800.}

\lref\rlenisch{W. Lerche,
B.E.W. Nilsson and A.N. Schellekens, 
{\it Heterotic String Loop Calculation of the Anomaly Cancelling Term},
Nucl.\ Phys.\ {\bf B289} (1987) 609.}
 
\lref\rkarch{A. Karch, D. L\"ust and D. Smith,
{\it Equivalence of Geometric Engineering and Hanany-Witten via Fractional
Branes}, Nucl.\ Phys.\ {\bf B533} (1998) 348, hep-th/9803232.}

\lref\randreas{B. Andreas, G. Curio and D. L\"ust,
{\it The Neveu-Schwarz Five-brane and its Dual Geometries},
JHEP {\bf 9810} (1998) 022, hep-th/9807008.}

\lref\rhanwit{A. Hanany and E. Witten,
{\it Type IIB Superstrings, BPS Monopoles and Three-dimensional Gauge Dynamics},
Nucl.\ Phys.\ {\bf B492} (1997) 152, hep-th/9611230.}

\lref\rdougmoor{M.R. Douglas and G. Moore,
{\it D-branes, Quivers and ALE Instantons},
hep-th/9603167.}

\lref\rmalda{J. M. Maldacena, {\it The Large N Limit of Superconformal 
Field Theories and Supergravity}, Adv.Theor.Math.Phys. 2 (1998) 231,
hep-th/9711200.}
 
\lref\rnekr{N. Nekrasov and S.L. Shatashvili, {\it On Non-Supersymmetric
    CFT in Four Dimensions}, hep-th/9902110.}

\lref\rpoly{A.M. Polyakov, {\it The Wall of the Cave}, 
Int.J.Mod.Phys. {\bf A14} (1999) 645, hep-th/9809057.}

\lref\rlnv{A. Lawrence, N. Nekrasov and C. Vafa, {\it On Conformal Field
Theories in Four Dimensions}, Nucl.\ Phys.\ {\bf B533} (1998) 199,
hep-th/9803076.}

\lref\rsilver{S. Kachru and E. Silverstein, {\it 4d Conformal Field Theories
and Strings on Orbifolds}, Phys. Rev. Lett. 80 (1998) 4855, hep-th/9802183.}

\lref\rkol{A. Armoni and B. Kol, {\it Non-Supersymmetric Large N Gauge
Theories from Type 0 Brane Configurations}, hep-th/9906081.}

\lref\rhanany{A. Hanany and E. Witten, {\it TypeIIB Superstrings, BPS 
monopoles and Three Dimensional Gauge Dynamics}, 
Nucl.\ Phys.\ {\bf B492} (1997) 152, hep-th/9611230.}

\lref\rwitten{E. Witten, {\it Solutions of Four-Dimensional Field Theories
Via M-Theory}, Nucl.\ Phys.\ {\bf B500} (1997) 3, hep-th/9703166.}

\lref\ranbadu{I. Antoniadis, C. Bachas and E. Dudas,
{\it Gauge Couplings in Four-dimensional Type I String Orbifolds},
hep-th/9906039.}

\lref\rbillo{M. Billo, B. Craps and F. Roose, {\it On D-branes in
Type 0 String Theory}, hep-th/9902196.}

\lref\rbfl{R. Blumenhagen, A. Font and D. L\"ust, {\it Tachyon-free
Orientifolds of Type 0B Strings in Various Dimensions}, hep-th/9904069.}

\lref\rkletsya{I.R. Klebanov and A.A. Tseytlin, {\it D-Branes and Dual Gauge 
theories in Type 0 Strings}, Nucl.Phys. {\bf B546} (1999) 155, 
hep-th/9811035.} 

\lref\rkletsyb{I.R. Klebanov and A.A. Tseytlin, {\it A Non-supersymmetric
Large N CFT from Type 0 String Theory}, JHEP {\bf 9903} (1999) 015,
hep-th/9901101.} 

\lref\rkletsyc{I.R. Klebanov and A.A. Tseytlin, {\it Asymptotic Freedom
and Infrared Behavior in the Type 0 String Approach to
Gauge Theory},  Nucl.\ Phys.\ {\bf B547} (1999) 143, hep-th/9812089}

\lref\rbergab{O. Bergman and M.R. Gaberdiel, {\it A Non-Supersymmetric Open 
String Theory and S-Duality}, Nucl.Phys. {\bf B499} (1997) 183, 
hep-th/9701137.}

\lref\rbergman{O. Bergman and M.R. Gaberdiel, {\it Dualities of Type 0
Strings}, hep-th/9906055.}

\lref\rangel{C. Angelantonj, {\it Non-Tachyonic Open Descendants of the 
0B String Theory}, Phys.Lett. {\bf B444} (1998) 309, hep-th/9810214.}

\lref\rbk{R. Blumenhagen and A. Kumar, {\it A Note on Orientifolds and 
Dualities of Type 0B String Theory}, hep-th/9906234.}

\lref\rfoe{K. F\"orger, {\it
On Non-tachyonic $Z_N\times Z_M$ Orientifolds of Type 0B String Theory},
hep-th/9909010.}

\lref\rsagbi{A. Sagnotti, M. Bianchi,
{\it On the Systematics of Open String Theories},
Phys.\ Lett.\ {\bf B247} (1990) 517}

\lref\rsagn{A. Sagnotti, {\it Some Properties of Open String Theories},
 hep-th/95090808 \semi
A. Sagnotti, {\it Surprises in Open String Perturbation Theory},
hep-th/9702093.}

\lref\rklewit{I.R. Klebanov and E. Witten, {\it Superconformal field 
theory on three-branes at a Calabi-Yau singularity}, 
Nucl. Phys. {\bf B536} (1998) 199, hep-th/9807080.}

\lref\rdixhar{L. Dixon and J. Harvey, {\it String Theories in Ten Dimensions
Without Space-Time Supersymmetry}, Nucl.\ Phys.\ {\bf B274} (1986) 93.}

\lref\rseibwit{
N. Seiberg and E. Witten, {\it Spin Structures in String Theory}, 
Nucl.\ Phys.\ {\bf B276} (1986) 272.}

\lref\roz{ M. Alishahiha, A. Brandhuber and Y. Oz, 
{\it Branes at Singularities in Type 0 String Theory}, 
     JHEP {\bf 9905} (1999) 024, hep-th/9903186.}

\Title{\vbox{
 \hbox{HUB--EP--99/56}\hbox{LPTENS-99/35}}}
 {\vbox
 {Continuous Gauge and Supersymmetry Breaking
\vskip0.2cm
\centerline{for Open Strings on D-branes}}
}
{\bf \centerline{Ralph Blumenhagen${}^1$, Costas Kounnas${}^{2,3}$ 
and Dieter L\"ust${}^4$}}
\bigskip
\centerline{\it ${}^{1,4}$ Humboldt-Universit\"at Berlin, Institut f\"ur 
Physik, Invalidenstrasse 110,}
\centerline{\it  10115 Berlin, Germany }
\smallskip
\centerline{\it ${}^2$ Laboratoire de Physique Th\'eorique, ENS,
F-75231 Paris, France}
\smallskip
\centerline{\it ${}^3$ Theory Divison, CERN, CH-1211,
Geneva 23, Switzerland}
\bigskip
\centerline{\bf Abstract}
\noindent
We consider  freely acting orbifold compactifications, which interpolate in 
two possible decompactification limits between the supersymmetric type II
string and the non-supersymmetric type 0 string. In particular we discuss
how D-branes are incorporated into these orbifold models. Investigating the 
open string spectrum on D3-branes, we will show that one can interpolate
in this way between  ${\cal N}=4$ supersymmetric 
$U(N)$ respectively  $U(2N)$ Yang-Mills theories and  non-supersymmetric
$U(N)\times U(N)$ gauge theories with adjoint massless
scalar fields plus bifundamental massless fermions
in a smooth way. Finally, by lifting the orbifold construction to
M-theory, we conjecture  some duality relations and show that in particular
a new supersymmetric branch of gauge like theories emanate 
for the  non-supersymmetric model.

\footnote{}
{\pano
${}^1$ e--mail:\ blumenha@physik.hu-berlin.de
\pano
${}^{3}$ e--mail:\ costas.kounnas@cern.ch
\pano
${}^4$ e--mail:\ luest@physik.hu-berlin.de
\pano}
\Date{10/99}


\newsec{Introduction}

As it became clear during the recent years, the emergence of 
$(p+1)$-dimensional,
supersymmetric
gauge theories from open strings living on the world volumes
of Dp-branes in closed string theories
provides very important insights into the perturbative and non-perturbative
gauge theory dynamics. Of particular interest is the limit in which
the closed (bulk) string degrees of freedom  decouple
from the open (boundary) degrees of freedom; this limit is provided
on the $U(N)$ gauge theory side by taking the number of color 
degrees of freedom
to be very large, $N\rightarrow\infty$, whereas on the closed string side
the same limit is achieved by considering the classical weak coupling
limit, $g_s\rightarrow 0$. 
Quite surprisingly, there is in fact a very striking duality between
the open string gauge theories on the boundary and the 
closed string gravitational theories in the bulk, stating that
superconformal gauge theories in the large N limit are dual to type II
superstring theories in anti-de-Sitter background 
spaces \refs{\rmalda\gkp-\wittena}.
The most well understood example is given by the duality
between ${\cal N}=4$, $U(N)$ Yang-Mills gauge theory and supergravity
in an $AdS_5\times S^5$ background.

Of course the most interesting problem is to explore the CFT/AdS
duality for models which allow for (spontaneous) breaking of space-time
supersymmetry. One way to break supersymmetry in four
dimensions is to consider models with non-zero temperature \wittenb.
A different and perhaps more direct way to construct non-supersymmetric
string models is provided by the type 0 string 
construction \refs{\rkletsya\rkletsyb-\rkletsyc} 
following an idea by Polyakov \rpoly.
Seen from a world sheet point a view the 
non-supersymmetric type 0A/B theories are simply obtained
by a non-supersymmetric, but nevertheless modular invariant GSO 
projection \refs{\rdixhar,\rseibwit},
such that the spectrum of closed strings is purely bosonic.
Alternatively the type 0 strings can be constructed as an orbifold of
type II superstring theory \rnekr\ by modding out the symmetry 
$(-1)^{F_s}$, where $F_s$ denotes the space-time fermion number.
In this way all closed string fermions in 
the (R,NS) and (NS,R) are projected out by the $\ZZ_2$ action, whereas
the twisted sector contains an (NS,NS) tachyon and in addition a second set of
massless (R,R) fields.
This has the important effect that the D-branes present
in type 0 theories
are essentially doubled compared to the supersymmetric type II parent
theories. We will call the two different kinds of p-dimensional D-branes
(electric) Dp and (magnetic) Dp'-branes.
Implementing the $(-1)^{F_s}$ projection 
in the open string sector, it turns out that the open
strings stretched between D-branes of the same kind, i.e. between
Dp and Dp-branes respectively between Dp' and Dp'-branes, lead to space-time
bosons, in particular to massless gauge bosons and massless scalar fields.
On the other hand, open strings between Dp and Dp'-branes lead to
space-time fermions. Let us mention that recently also 
non-supersymmetric tachyon-free compactifications of both type II 
string theory
\refs{\rkks\rharv\rshiutye\rblgo-\raaf} and type 0 string theory
\refs{\rsagbi\rsagn\rbergab\rangel\rbfl\rbk-\rfoe} were discussed. 

Using this construction an  interesting class of non-supersymmetric gauge 
models arises when considering an equal number of $N$ electric D3 and $N$ 
magnetic D3'-branes in type 0B
string theory. This configuration leads to a non-supersymmetric
$U(N)_e\times U(N)_m$ gauge theory with bosonic and fermionic massless
matter arranged in  such a way 
that the $\beta$-function vanishes to leading order
in $1/N$. In fact, Klebanov and Tseytlin 
\refs{\rkletsya\rkletsyb-\rkletsyc}
gave reasonable evidence
that this gauge theory is again dual to a tachyon-free gravity background
of the form $AdS_5\times S^5$ just as in the supersymmetric type IIB
theory. Putting the type 0B D3-branes on transversal 
singularities \refs{\rdougmoor\rsilver\rlnv-\rklewit}
or considering type 0A Hanany-Witten \refs{\rhanany} 
like brane constructions a full variety
of non-supersymmetric gauge theories $(U(N)_e\times U(N)_m)^K$
can be constructed
\refs{\roz\rbillo\rkol\rbfla-\kns}, 
where $K$ is a number which characterizes the singularity
type or the number of NS 5-branes in the Hanany-Witten set up. All these
models are Bose-Fermi degenerated and have the same one-loop
$\beta$-functions as their corresponding 
$U(N)^K$ type II parent models. 
This correspondence becomes even more close by noting 
that the
gauge theory given by the diagonal $U(N)_{e+m}\in U(N)_e\times U(N)_m$
is identical to the gauge theory of the underlying type II theory.
Therefore one might conjecture that there exists a smooth interpolation 
between the supersymmetric
$U(N)$ gauge theories and the non-supersymmetric $U(N)\times U(N)$
models in such a way that supersymmetry is smoothly broken (installed)
along the line of deformation. 
Since the $U(N)_e\times U(N)_m$ gauge theories do not contain any
massless scalars fields with the right quantum numbers to trigger
the symmetry breaking to the diagonal group $U(N)_{e+m}$ via the usual
field theoretical Higgs mechanism, one has to consider the couplings between
the open string boundary modes to the closed string bulk modes
in order to realize the desired interpolation.

One well established way to break supersymmetry in
a smooth way is provided by the Scherk-Schwarz 
mechanism \refs{\scherk\rKP\fkpz\rkonrost\koubps-\KK}.
This class of models is built as freely acting orbifolds with different
boundary conditions for bosons and fermions, very similar to strings
at finite temperature \refs{\sat\kog\ratwitt\rkonrost\rAK\ADKa-\ADKb}.
Sending the radius $R$ of the orbifold to infinity (to zero) space-time
supersymmetry is recovered.
Recently Scherk-Schwarz  supersymmetry breaking was
studied in type I compactifications \refs{\rdienes\adsa\aads-\ads}
and also in M-theory \aq. 
Moreover the Scherk-Schwarz mechanism was also used  to show that there
exist indeed two types of models which smoothly interpolate between 
the closed
type IIA/B and type 0A/B theories \rbergman. They can be either
realized as a freely acting orbifold of type II on $S^1/(-1)^{F_s}S$, where
$S$ denotes a half shift on the circle, or alternatively as a freely acting
orbifold of type 0 on $S^1/(-1)^{f_L}S$, where $f_L$ is the left-moving
world sheet fermion number. In the former case supersymmetry is
recovered for infinite radius of the circle, whereas in the 
latter case the zero radius
limit is fully supersymmetric. These interpolating models are expected to
exhibit a Hagedorn
phase transition at the special critical radius of the circle where
a tachyonic mode arises \refs{\ratwitt\rkonrost\rAK\ADKa-\ADKb}.
The freely acting orbifold of type IIA   can also be  lifted 
to M-theory \rbergman. 
In this case it was 
postulated in \rbergman\  that the strong coupling regime of type 0A 
is given by  M-theory on $S^1/(-1)^{F_s}S$. Taking this conjecture seriously,
one implication is that at strong coupling the type 0A string
becomes supersymmetric in the bulk. Moreover, the closed string tachyon 
becomes
massive at strong coupling and at the same time, fermions, which are
of solitonic nature in the type 0A string, 
become light and provide the massless
(R-NS) fermions of the closed type IIA superstring.

In this paper we will extend this construction including also the
D-branes and the corresponding open strings into the orbifolds which
interpolate between type II and type 0 strings.
In this way we will show that one can smoothly
interpolate between 
broken and unbroken gauge theories and
at the same time between supersymmetric and non-supersymmetric 
models in a way not known before from field theory.
We will first consider type IIA/B compactified on $S^1/(-1)^{F_s}S$
where $N$ Dp-branes (p even (odd) for type A(B)) are either wrapped around
the orbifold circle, or are placed 
transversal to it. In the former case one interpolates 
between a $(p+1)$-dimensional, ${\cal N}=4$ supersymmetric $U(2N)$ 
gauge theory at infinite radius
and a $p$-dimensional, non-supersymmetric $U(N)\times U(N)$ gauge at zero
radius. In case of transversal branes, the infinite radius limit of the
orbifold provides an ${\cal N}=4$ supersymmetric $U(N)$ gauge theory in $p+1$
dimensions whereas  in the zero radius limit a non-supersymmetric
$U(N)\times U(N)$ gauge theory in $p+2$ dimensions appears.
We extend this construction by considering two-dimensional
type IIA/B orbifold
compactifications on $S^1\otimes S^1/(-1)^{F_s}S$. In turns out that
varying
the two radii of the compact space one can 
interpolate between supersymmetric and non-supersymmetric gauge theories
living in the same space-time dimension. 
A similar, in fact T-dual, picture arises from D-branes of type 0A/B 
on $S^1/(-1)^{f_L}S$
and on $S^1\otimes S^1/(-1)^{f_L}S$, respectively.

Finally we will discuss how this construction can be embedded
into M-theory. Here again the M-theory moduli allow  
to interpolate between weakly coupled non-supersymmetric
gauge theories and strongly coupled supersymmetric gauge theories in a smooth
way.
In contrast to the string theory embeddings, here one of the radii is related
to the gauge coupling constant, so that one is really interpolating between
small and strong gauge coupling. 
Taking the non-supersymmetric dualities seriously, we are led to
strong-weak coupling dualities for the non-supersymmetric gauge theories.
Moreover, we see 
that non-supersymmetric gauge theories contain new branches
at strong coupling, which are supersymmetric, but not all stringy degrees
of freedom are decoupled.

\newsec{Type II compactifications on freely acting orbifolds}

\subsec{Supersymmetry restoration in the bulk}

In this section 
we briefly review the smooth orbifold construction of \rbergman\ 
which produces
type 0 from type II and vice versa.
One starts with the type IIA (B) superstring compactified in the ninth
direction on a circle $S^1$ of radius $R_9^{II}$. 
The theory contains 32 unbroken supercharges for every value of
$R_9^{II}$. In the limit $R_9^{II}
\rightarrow \infty$
the Kalazu-Klein modes become massless and one recovers
the type IIA (B) superstring in ten dimensions. On the other hand, for 
$R_9^{II}\rightarrow 0$ the winding modes become 
massless and the theory is now 
identical
to the type IIB (A) superstring in ten dimensions.
This is what is usually called T-duality between the type IIA/B 
superstring \refs{\dhs,\dlp} (for a recent discussion on IIA/IIB T-duality
and M-theory see \adln).

At the next step we build the orbifold type IIA (B) on
$S^1/(-1)^{F_s}$ of radius $R_9^0$. This $\ZZ_2$ projection breaks all
32 supercharges and the resulting theory is nothing else than
the type 0 string compactified on $S^1$ with purely bosonic spectrum in the 
closed string bulk. 
For large $R_9^0\rightarrow\infty$ 
the type 0A (B) string in ten dimensions emerges whereas for 
$R_9^0\rightarrow 0$ the ten-dimensional type 0B (A) string is present.
Hence $T$-duality among type 0A/B just works like for the 
type IIA/B superstring pair.

Finally we combine the action of the $\ZZ_2$ $(-1)^{F_s}$ orbifold projection
with the half-shift $S$ along  the circle, i.e. we are now considering
type IIA (B) on the freely acting orbifold $S^1/(-1)^{F_s}S$ of radius $R_9$.
For finite values of $R_9$ all 32 supercharges are broken in the sense
that all 32 gravitinos are massive with mass of order $1/R_9$.
In the limit $R_9\rightarrow\infty$ the gravitinos
become massless, and one obtains the closed
string spectrum of the ten-dimensional type IIA (B) superstring.
On the other hand, 
for $R_9\rightarrow 0$ the gravitinos become infinitely heavy such that they
completely decouple, and one regains the spectrum  of the type 0A (B)
for $R_9^0$=0. So effectively, using also the  type 0 T-duality,
the limit $R_9\rightarrow 0$ of type IIA (B) on this orbifold is given
by the ten-dimensional  type 0B (A) string theory.  
There is a tachyon for $R_9<\sqrt{2}$, which  decouples in
the limit $R_9\rightarrow\infty$.
Therefore one expects a Hagedorn phase transition at this 
radius \refs{\rAK\ADKa-\ADKb}.
Note that at infinite radius the two moduli spaces of  type II 
compactified on the circle $S^1$ and of type II compactified
on the orbifold $S^1/(-1)^{F_s}S$ meet. In the same way, at $R_9=0$ the
moduli space of type 0 on $S^1$ and of type II on $S^1/(-1)^{F_s}S$
have a common intersection.

\subsec{D-branes and open strings in Type II on $S^1/(-1)^{F_S}S$}

First let us briefly recall the D-brane spectrum of type 0 strings. 
As already mentioned, the $(-1)^{F_s}$ projection removes
all closed string 
fermions from the untwisted sector and leads to a tachyon and further
massless RR fields in the twisted sector. 
Computing the spectrum one finds that all states in the RR sector are 
doubled implying that all Dp-branes (p odd in type 0B, p even
in type 0A) are doubled, as well. 
In particular,  in the case of D3-branes 
there now exist (electric) D3 and (magnetic) D$3'$-branes.
Using the boundary state approach it was shown in \rbergab\ that indeed the 
boundary
state representing a Dp-brane in type II splits into two boundary states
of type 0. The explicit form of the boundary states was used to derive
rules for open strings stretched between the various types of Dp-branes.
Open strings stretched between the same kind of Dp-branes carry only
space-time bosonic modes, whereas open strings stretched between a Dp and
a D$p'$-brane carry only space-time fermionic modes.
The massless spectrum living on the D3-branes is given by a four-dimensional 
gauge theory with
gauge group $G={U(N)}\times {U(N)}$,
          three complex bosons in the
           adjoint and four Weyl-fermions in the $(\rN,\o{\rN})+(\o{\rN},\rN)$
           representation of $G$.
It follows that 
the one-loop $\beta$-function vanishes, and as was shown explicitly
in \rkletsyb\  the two-loop $\beta$-function vanishes in the large 
N limit. In the next to leading order one obtains a non-vanishing 
contribution to the two-loop $\beta$-function coefficient, $b_2=-16$. 
Thus, the gauge theory is free in the infrared. 

Next consider the type II string
compactified on $S^1/(-1)^{F_s}$.
First we discuss the case where the circle $S^1$ is transversal to the
Dp and Dp'-branes such that the D-branes take certain positions on $S^1$.
In addition to the open strings discussed before there are also winding
open strings which are wrapped $n$-times around $S^1$ before their ends
are stuck at the D-branes. 
Therefore we call this case the {\it winding picture}.
The masses of these winding states are
proportional to $|nR_9^0|$. 
On the one hand, 
for $R_9^0\rightarrow \infty$ the winding modes become
heavy and decouple.
On the other hand for $R_9^0\rightarrow 0$
an infinite number of winding modes becomes massless, which means that
effectively the world volume dimension of a Dp-brane grows by one unit.

Alternatively we can also decide to put the compact circle along one of the
world volume directions of the Dp-branes. This is called the {\it momentum
picture}, since there are open strings on the Dp(p')-branes 
which carry discrete
momenta
of the order $m/R_9^0$.
Hence these states become light in the limit $R_9^0\rightarrow\infty$, but
decouple in the other decompactification limit $R_9^0\rightarrow 0$.
Thus, in the $R_9^0\rightarrow \infty$  limit the dimension of the brane 
world-volume is effectively enlarged.

After these considerations we are now ready to discuss the
D-branes and open strings within the compactification
of type II string on the orbifold $S^1/(-1)^{F_s}S$.

\vskip0.3cm

\noindent {\it (i) The winding picture: transversal Dp-branes}

\vskip0.3cm

For concreteness let us consider the 
type IIB string compactified on a circle of radius
$R_9$. We place 2N transversal D3-branes in this background
and divide by the $\ZZ_2$ operation $T=(-1)^{F_s} S$.
The shift acts on a momentum/winding
ground state as $S|m,n\ra= (-1)^m |m,n\ra$. Let us first consider
the range $0<R_9<\infty$ and then consider the two possible limits.
Requiring that $T$ is indeed a symmetry of the background, we have to
make sure that the D3-branes are arranged in such a way that $T$ transforms
one brane into another. We will place N branes at a position $x_9=A=0$
on the circle and N branes at $x_9=B=R_9/2$. 
Therefore we deal with two kinds of winding strings in the open
string sector. First there are the AA- and BB-sectors
with open strings between two D3-branes either at $x_9=0$
or at $x_9=R_9/2$ and integer winding numbers
$w=nR_9$. Their masses
are given by
\eqn\massw{
M^2\sim(nR_9)^2.} 
Second in the AB- and
BA-sectors there are open strings stretched 
between one D3-brane at $x_9=0$ and one other D3-brane at $x_9=R_9/2$.
These states have half-inter winding numbers $w=(n+\frac{1}{2})R_9$, and
their masses 
are given by 
\eqn\massm
{M^2\sim((n+\frac{1}{2})R_9)^2.} 
Hence the ground
state  in the AB(BA)-sector is massive for finite $R_9$.

The natural action
of $T$ on the Chan-Paton factors of these branes is given by
\eqn\chapaa{  \gamma_{T}= 
                        \left(\matrix{  0  &  I   \cr
                                        I  &  0  \cr} \right)_{2N,2N} }
so that the branes at $x_9=0$ are mapped to the branes at $x_9=R_9/2$.
Computing the annulus amplitude for open string stretched between
two D3-branes is straightforward and gives 
\eqn\anna{\eqalign{ A=&{\rm Tr}\left[ {1+T\over 2}\, P_{GSO}\, 
                e^{-2\pi t L_0} \right] \cr
                  =&{N^2\over 2}\ 
      {\th{0}{0}^4-\th{0}{1/2}^4-\th{1/2}{0}^4 \over \eta^{12} }
               \sum_{n\in\ZZ} \left( e^{-2\pi t R_9^2 n^2} +
                          e^{-2\pi t R_9^2 (n+{1\over 2})^2} \right), }}
with argument $e^{-2\pi t}$. 
The first term with integer windings is from open strings stretched between
branes at the same location and the second term with half-integer windings
is from open strings stretched between branes at opposite locations of the 
circle. 
It is straightforward to compute the massless spectrum. In the range 
$R_9>0$ we obtain the four dimensional supersymmetric spectrum 
shown in Table 1.
\vskip0.5cm
\meno
\cl{\vbox{
\hbox{\vbox{\offinterlineskip
\def\tablespace{height2pt&\omit&&\omit&&
 \omit&\cr}
\def\tablerule{\tablespace\noalign{\hrule}\tablespace}

\hrule\halign{&\vrule#&\strut\hskip0.2cm\hfil#\hfill\hskip0.2cm\cr
\tablespace
& sector && spin && gauge    $U(N)$  &\cr
\tablerule
& AA, BB   && vector && {\bf Adj} &\cr
\tablespace
&   && scalar && $6\times  {\bf Adj} $ &\cr
\tablespace
&   && Weyl && $4\times  {\bf Adj} $ &\cr
\tablespace}\hrule}}}}
\cl{
\hbox{{\bf Table 1:}{\it ~~ Winding picture: 
open string spectrum of $S^1/T$ for finite $R_9$}}}
\smno
\vskip0.5cm\noindent 
Thus the massless spectrum 
in the AA and BB sectors is exactly the one of $U(N)$ ${\cal N}=4$
super Yang-Mills theory in four space-time 
dimensions. 
In the limit
$R_9\to\infty$ the mass of the other winding states becomes infinite and 
they decouple.
Therefore we are dealing with four-dimensional
$U(N)$ ${\cal N}=4$ supersymmetric Yang-Mills in this decompactification 
limit.

In the $R_9\to 0$ limit however 
an infinite number of winding states  in the AA and BB sectors becomes 
massless.
This has the effect that the gauge theory now lives in one dimension
higher. Moreover
the gauge group gets enlarged, 
as now open strings in the AB and BA sectors
stretched between the D3-branes at locations
A and B become massless. By a unitary transformation
\eqn\chapab{ U = {1\over\sqrt{2}}
                        \left(\matrix{  I  &  I   \cr
                                        I  &  -I  \cr} \right)_{2N,2N} }
one obtains $\gamma_T={\rm diag}[I,-I]$ and now one is
in the situation of the
pure $(-1)^{F_s}$ orbifold, which leads to the massless open string
spectrum listed in Table 2.
\vskip0.5cm
\meno
\cl{\vbox{
\hbox{\vbox{\offinterlineskip
\def\tablespace{height2pt&\omit&&\omit&&
 \omit&\cr}
\def\tablerule{\tablespace\noalign{\hrule}\tablespace}

\hrule\halign{&\vrule#&\strut\hskip0.2cm\hfil#\hfill\hskip0.2cm\cr
\tablespace
& sector && spin && gauge    $U(N)\times U(N)$  &\cr
\tablerule
& 33, 3'3'   && vector && ({\bf Adj},1)+(1,{\bf Adj}) &\cr
\tablespace
&   && scalar && $5\times  \{ ({\bf Adj},1) + (1, {\bf Adj}) \} $ &\cr
\tablerule
& 33', 3'3  && Dirac && $2\times\{ (N,\o{N}) + (\o{N},N) \} $ &\cr
\tablespace}\hrule}}}}
\cl{
\hbox{{\bf Table 2:}{\it ~~ Winding picture: 
open string spectrum of $S^1/T$ for $R_9=0$ }}} 
\smno
\vskip0.5cm
This is precisely the mass spectrum of type 0B with $N$ electric D3 plus $N$
magnetic D3'-branes but at zero radius $R_9^{0B}=0$.
This means that the non-supersymmetric gauge theory actually lives in
five space-time dimensions, i.e. this limit is nothing else than the type 0A
string with a non-supersymmetric 
$U(N)\times U(N)$ gauge theory arising from $N$ electric D4
plus $N$ magnetic D4'-branes.

In summary, 
so far we have constructed an interpolating model between 
four-dimensional ${\cal N}=4$ super YM theory
at $R_9\to \infty$ and a special five-dimensional non-supersymmetric 
gauge theory at
$R_9\to 0$. 
The additional massless states
appear in the winding sector of open strings; hence there
is no field theoretic description for them.
$R_9$ is a purely stringy parameter, which  is not
present  in conventional  gauge theory. 
The corresponding closed string modulus
field in the bulk is coupled to the open strings in the boundary
and provides
at the same time the supersymmetry restoration and the gauge symmetry breaking
on the D-branes.

Of course, this picture can be immediately generalized considering
2N Dp-branes, p even (odd), in type IIA(B) superstring compactified
on $S^1/(-1)^{F_s}S$, where again half of the Dp-branes are positioned
at $x_9=0$ and  the other half sit at $x_0=R_9/2$.
In this way one interpolates between  supersymmetric type IIA(B)
Dp-branes at $R_9\rightarrow\infty$ and non-supersymmetric
type 0B(A) D(p+1)-, D(p+1)'-branes at $R_9\rightarrow 0$.
In the open string sector one is smoothly interpolating
between (p+1)-dimensional supersymmetric $U(N)$ Yang-Mills theory with
16 supercharges at
$R_9=\infty$ and (p+2)-dimensional non-supersymmetric $U(N)\times U(N)$
Yang-Mills theory at $R_9=0$.

\vskip0.3cm
\vfill\eject

\noindent {\it (ii) The momentum picture: wrapped Dp-branes}

\vskip0.3cm

Now let us discuss the case where the orbifold $S^1/(-1)^{F_s}S$ lies
in one of the world volume directions of the D-branes, i.e. the
D-branes are wrapped around the compact orbifold.
The open string states again split into two sectors, namely momentum states
with even momenta $p_9=2m/R_9$ 
and masses
\eqn\momass
{M^2\sim (2m)^2/R_9^2,}
and  in general different states with
odd momenta $p_9=(2m+1)/R_9$ and corresponding masses
\eqn\momassa
{M^2\sim (2m+1)^2/R_9^2.}
To be specific consider 2N D4-branes of type IIA wrapped in this way.
For finite values of $R_9$ the resulting gauge theory lives in four
uncompactified space-time dimensions, whereas in the  $R_9\rightarrow
\infty$ limit momentum modes are massless, and 
the gauge theory become five-dimensional.
Since in the $R_9\to 0$ limit we would like to obtain the non-supersymmetric
gauge theory with gauge group $U(N)\times U(N)$ we make the following choice 
for the action of $T$ on the Chan-Paton labels
\eqn\chapab{  \gamma_{T}= 
                        \left(\matrix{  I  &  0   \cr
                                        0  &  -I  \cr} \right)_{2N,2N} .}
In order to obtain the precise form of the spectrum we  compute
the annulus diagram for open strings stretched between the wrapped
D4-branes. 
\eqn\annb{\eqalign{ A=&{\rm Tr}\left[ {1+T\over 2}\, P_{GSO}\, 
                e^{-2\pi t L_0} \right] 
                  ={N^2}\ {\th{0}{0}^4-\th{0}{1/2}^4-\th{1/2}{0}^4 \over
        \eta^{12}  }\
               \sum_{m\in\ZZ} \left( e^{-2\pi t {m^2\over R_9^2}} \right). }}
The resulting massless spectrum for $R_9<\infty$ is listed in Table 3.
\vskip0.5cm
\meno
\cl{\vbox{
\hbox{\vbox{\offinterlineskip
\def\tablespace{height2pt&\omit&&\omit&&
 \omit&\cr}
\def\tablerule{\tablespace\noalign{\hrule}\tablespace}

\hrule\halign{&\vrule#&\strut\hskip0.2cm\hfil#\hfill\hskip0.2cm\cr
\tablespace
& sector && spin && gauge  $  U(N)\times U(N)$  &\cr
\tablerule
&  && vector && $({\bf Adj},1)+(1,\bf {Adj})$ &\cr
\tablespace
& $p_9$ even  && scalar && $6\times \{ ({\bf Adj},1)+
                    (1,{\bf Adj})\} $ 
&\cr
\tablespace
&   && Weyl && $4\times  \{ (N,\o{N}) + (\o{N},N) \}   $ &\cr
\tablespace}\hrule}}}}
\cl{
\hbox{{\bf Table 3:}{\it ~~ Momentum picture: 
open string spectrum of $S^1/T$ for finite $R_9$}}}
\smno
\vskip0.5cm
We see indeed that the massless spectrum is the one of four-dimensional,
non-supersymmetric Yang-Mills with gauge group $U(N)\times U(N)$. The
massless spectrum agrees with the open string spectrum of D3- and D3'-branes
in the type 0B string. Therefore in the limit $R_9\rightarrow 0$ 
the massive states with masses proportional to $(2m+1)/R_9$ decouple, and we
get precise agreement with the type 0B string.

The limit $R_9\rightarrow
\infty$ provides on the other 
hand an infinite number of new massless momentum states
which enhance the gauge symmetry to the group
$U(2N)$ and also restores space-time
supersymmetry. So in the large radius limit
the spectrum is given by supersymmetric $U(2N)$ Yang-Mills theory in five
space-time dimensions, where 16 supercharges are preserved in
the open string theory.

In general starting with $2N$ wrapped Dp-branes in type IIA(B)
on $S^1/(-1)^{F_s}S$ one interpolates between the following
two decompactification limits:

\noindent 
$R_9\rightarrow 0$: N D(p-1) and N D(p-1)'-branes of type 0B(A)
with p-dimensional, non-supersymmetric $U(N)\times U(N)$ gauge theory.

\noindent $R_9\rightarrow\infty$: 2N Dp-branes of type IIA(B) with
(p+1)-dimensional, supersymmetric $U(2N)$ gauge theory.

\subsec{D-branes and open strings in type II on $S^1\otimes S^1/(-1)^{F_S}S$}

So far we have interpolated between supersymmetric and non-supersymmetric
gauge theories living in different number of space-time dimensions.
Now we will extend the discussion by compactifying the type II
string on a two-dimensional compact space given
by $S^1\otimes (S^1/(-1)^{F_s}S)$ characterized by the two 
radii $R_8$ and $R_9$. 
Using the T-duality on the compact circle in $x_8$ we can
now smoothly interpolate between D-branes of the same world volume
dimensions, and hence also between non-supersymmetric and supersymmetric
gauge theories of the same dimensionality.

\vskip0.3cm

\noindent {\it (i) The winding picture: transversal Dp-branes}

\vskip0.3cm

Let us consider 2N Dp-branes in type IIA(B)
which are transversal to both  the
$x_8$ and the $x_9$ direction. The positions of the Dp-branes on the
orbifold circle are as discussed in section (2.1).
The spectrum of this theory follows without large efforts from the previous
discussion. For finite values of the two radii $R_8$ and $R_9$ 
we are dealing with a (p+1)-dimensional gauge theory, living
on the world volume of the Dp-branes. The massless states are those
of maximally supersymmetric $U(N)$ Yang-Mills gauge theory.
Of particular interest are the following four possible decompactification
limits:

\vskip0.2cm
\noindent a) $R_8\rightarrow\infty$, $R_9\rightarrow\infty$: 
Here we are dealing with Dp-branes of type IIA(B), and the corresponding
gauge theory is (p+1)-dimensional, supersymmetric Yang-Mills
with $U(N)$ gauge group.

\vskip0.2cm
\noindent b) $R_8\rightarrow 0$, $R_9\rightarrow\infty$:
This limit describes D(p+1)-branes of type IIB(A) with (p+2)-dimensional,
supersymmetric $U(N)$ gauge theory.

\vskip0.2cm
\noindent c) $R_8\rightarrow\infty$, $R_9\rightarrow 0$:
Now the decompactification limit corresponds to self-dual D(p+1)- and
D(p+1)'-branes of type 0B(A) with non-supersymmetric
$U(N)\times U(N)$ gauge theory in p+2 dimensions.

\vskip0.2cm
\noindent d) $R_8\rightarrow 0$, $R_9\rightarrow 0$:
Finally we obtain self-dual D(p+2), D(p+2)'-branes of type 0A(B)
with non-supersymmetric $U(N)\times U(N)$ gauge theory in p+3 dimensions.

\vskip0.2cm

Therefore, in the winding picture the moduli space of this two-dimensional 
compactification enables us to interpolate between N D3-branes of type IIB 
and N self-dual D3, D3'-branes of type 0B.
Hence there exists a stringy interpolation mechanism between 
four-dimensional, non-supersymmetric $U(N)\times U(N)$ gauge theory
with 6 adjoint scalars plus $4\times \{ (N,\o{N})+(\o{N},N)\}$ Weyl fermions
and four-dimensional ${\cal N}=4$ supersymmetric $U(N)$ Yang-Mills theory.
The supersymmetric $U(N)$ gauge group is just the diagonal subgroup
of the non-supersymmetric $U(N)\times U(N)$ gauge symmetry.

\vskip0.3cm

\noindent {\it (ii) The momentum picture: wrapped Dp-branes}

\vskip0.3cm

In the momentum picture we are considering 2N Dp-branes which are
wrapped both in the $x_8$ and also in the $x_9$ directions.
For finite values of the radii there is a non-supersymmetric
$U(N)\times U(N)$ gauge theory with massless fields living in
p-1 uncompactified space-time dimensions.
Again we like to consider the four special decompactification limits:

\vskip0.2cm
\noindent a) $R_8\rightarrow\infty$, $R_9\rightarrow\infty$:
Here we are dealing with Dp-branes of type IIA(B), and the corresponding
gauge theory is (p+1)-dimensional, supersymmetric Yang-Mills
with $U(2N)$ gauge group.

\vskip0.2cm
\noindent b) $R_8\rightarrow 0$, $R_9\rightarrow\infty$:
This limit describes D(p-1)-branes of type IIB(A) with p-dimensional,
supersymmetric $U(2N)$ gauge theory.

\vskip0.2cm
\noindent c) $R_8\rightarrow\infty$, $R_9\rightarrow 0$:
Now the decompactification limit corresponds to self-dual D(p-1)- and
D(p-1)'-branes of type 0B(A) with non-supersymmetric
$U(N)\times U(N)$ gauge theory in p dimensions.

\vskip0.2cm
\noindent d) $R_8\rightarrow 0$, $R_9\rightarrow 0$:
Finally we obtain self-dual D(p-2), D(p-2)'-branes of type 0A(B)
with non-supersymmetric $U(N)\times U(N)$ gauge theory in p-1 dimensions.

\vskip0.2cm

We see that within the momentum picture one can
again interpolate between  2N D3-branes of type IIB and 
N self-dual D3, D3'-branes of type 0B.
This time it provides an interpolation mechanism between
four-dimensional, non-supersymmetric $U(N)\times U(N)$ gauge theory
with 6 adjoint scalars plus $4\times \{ (N,\o{N})+(\o{N},N)\}$ Weyl fermions
and four-dimensional ${\cal N}=4$ supersymmetric $U(2N)$ Yang-Mills theory.
Here the non-supersymmetric $U(N) \times U(N)$ gauge group is a regular
subgroup of the supersymmetric $U(2N)$ gauge symmetry.

\newsec{Type 0 compactifications on freely acting orbifolds}

Alternatively we can start also with the type 0A (B) string compactified
on $S^1$. Modding
by $(-1)^{f_L}$ leads back to the type IIA (B) superstring theories.
It follows that the compactification of type 0A (B) on the freely acting
orbifold $S^1/(-1)^{f_L}S$ possesses the following decompactification limits.
For $R_9\rightarrow\infty$ one recovers the ten-dimensional
non-supersymmetric type 0A (B)
theories and for $R_9\rightarrow 0$ the ten-dimensional supersymmetric
type IIB (A) theory arises.
Observe  that now a tachyon develops   for large radii
$R_9>\sqrt{2}$. As was shown in \rbergman\ the type 0A (B) orbifold is T-dual 
to the type IIB (IIA) orbifolds discussed in the previous section, where the
relation between the radii is $R_{II}=2/R_{0}$. The straightforward T-dual
of the type IIB (IIA) over $S^1/(-1)^{F_S}S$ orbifold is of course
type IIA (IIB) over $S^1/(-1)^{F_S}\tilde S$, where $\tilde S$ is a shift
in the momentum lattice acting as $(-1)^n$ on winding modes.

Let us now discuss the properties of the D-branes and the gauge theories
in the open string sector within this orbifold compactification
of the type 0 string. Since the situation is just T-dual to the orbifold
compactification of the type II strings discussed before, 
we will only summarize the main results.
In order to interpolate between type 0 and type II we will only allow
for self-dual D-branes in the type 0 orbifolds, i.e. we are placing
N electric Dp and N magnetic Dp'-branes either transversal to the orbifold
or wrapped around it.

\vskip0.3cm

\noindent {\it (i) The winding picture: transversal Dp-branes}

\vskip0.3cm

Here we are placing N electric Dp and N magnetic Dp'-branes both at
$x_9=0$ and at $x_9=R_9/2$. The action of $(-1)^{f_L}S$ on the CP-factors
relates for instance a Dp-brane at $x_9=0$ to a Dp'-brane
at $x_9=R_9/2$. Using the T-duality  relation between the type II and type 0
radii, one obtains exactly the annulus partition function in \annb.
For generic radii the open string sector is given by (p+1)-dimensional,
non-supersym\-me\-tric
$U(N)\times U(N)$ gauge theory. For $R_9\rightarrow 0$ additional gauge bosons
as well as their superpartners become massless, such
that this limit is described by D(p+1)-branes with associated ${\cal N}=4$
supersymmetric $U(2N)$ gauge theory in p+2 space-time dimensions.
On the other hand, for $R_9\rightarrow\infty$ we obtain the Dp, Dp'-branes
of type 0, and the gauge theory is non-supersymmetric $U(N)\times U(N)$
Yang-Mills theory.
Of course, we can extend this picture by considering a type 0 compactification
on $S^1\otimes S^1/(-1)^{f_L}S$. In this way we get the T-dual picture
of the interpolating models discussed in section 2.3.

\vskip0.3cm

\noindent {\it (ii) The momentum picture: wrapped Dp-branes}

\vskip0.3cm

Here the situation is similar to the winding picture in the type II
orbifold compactification. We are wrapping N electric Dp and 
N magnetic Dp'-branes around the circle and using the T-duality relation
for the radii we  obtain the same result as in \anna.
For $R_9\rightarrow 0$ the theory becomes ${\cal N}=4$ supersymmetric,
but now with $U(N)$ gauge group. In contrast, 
for $R_9\rightarrow\infty$  supersymmetry is completely broken but
the gauge group is enhanced to $U(N)\times U(N)$.

\newsec{M-theory embedding}

\subsec{The bulk theory}

So far we have considered embeddings of supersymmetric and non-supersymmetric
gauge theories into string theory and by varying some
stringy parameters we have seen that one can interpolate between them.
Unfortunately, this does not teach us anything new about the dynamics
of the non-supersymmetric gauge theories. Therefore, we should try
to lift the whole picture to M-theory where at least one of
the radii is related to the string coupling constant and therefore to the
gauge coupling constant. 

Let us briefly review the suggestion of \rbergman\ 
where a freely acting orbifold
of 11-dimensional M-theory was constructed with the conjecture
that the non-supersymmetric type 0A string can be viewed as an M-theory
compactification on $S^1/(-1)^{F_s}S$. As a consequence of this conjecture
it follows that at small radius $r_{10}$, i.e. at weak string coupling,
one recovers the non-supersymmetric type 0A string, whereas at large radius
$r_{10}$, i.e. at strong string coupling, the maximally supersymmetric
M-theory in 11 dimensions emerges. Hence the type 0A string should
contain fermionic solitons with masses $m_f=1/r_{10}$, and the tachyon of type
0A should become massive at strong coupling.
Similarly it was argued in \rbergman\ that the ten-dimensional
type 0B string can be obtained as the zero volume limit of M-theory on
$T^2/(-1)^{F_s}S$.






So let us consider M-theory compactified on the freely acting
orbifold $S^1\otimes S^1/(-1)^{F_s}S$.
The corresponding two radii are called $r_9$ and $r_{10}$.${}^6$\footnote{}
{${}^6$ Radii measured in  units of the 11-dimensional Planck length $L_{11}$
are denoted by small letters. On
the other hand, radii measured in units of the string scale $\sqrt{\alpha '}$
are  denoted as before by capital letters.}
If the $\ZZ_2$ orbifold is in the $x_9$ direction then the previous
perturbative discussion on the interpolation between supersymmetric
and non-supersymmetric theories
applies. However if we exchange the role of the two circles such that
supersymmetry is broken by 
the M-theory coordinate $x_{10}$,
one has to conclude that the type 0A string at weak coupling
is given by 
M-theory on $S^1/(-1)^{F_s}S$ at zero radius $r_{10}$.
Varying the M-theory radius $r_{10}$, i.e. the 0A coupling constant
$g_{10}^A$, one is interpolating between  M-theory with 32 
supercharges and type 0A string with completely broken supersymmetry.
To be precise 
let us recall the well-known relations \refs{\wittenc,\rschw} 
between the string and M-theory parameters. 
First the relation between the string coupling
constant $g_{10}^A$ and $r_{10}$ is given by ${}^7$\footnote{}
{${}^7$ The relation between the type 0 parameters and M-theory
parameters are as in the type II case except some additional factors of 2
due to the orbifoldization.}
\eqn\mtwoa{
g_{10}^A=\biggl(\frac{r_{10}}{2}\biggr)^{3/2}.}
Next we consider the compactification of M-theory 
to nine dimensions.
The nine-dimensional radius of type 0A on $S^1$ is then related
to the M-theory parameters as
\eqn\ninertwoa{
R_9^A=r_9\sqrt{\frac{r_{10}}{2}},}
whereas the nine-dimensional 0A string coupling constant reads
\eqn\ninegtwoa{
g_9^A=\frac{g_{10}^A}{\sqrt{R_9^A}}=\biggl(\frac{r_{10}}{2}\biggr)^{5/4}
r_9^{-1/2}.}
Using the T-duality between 0A and 0B in nine dimensions we can express
the 0B parameters in the following way:
\eqn\twob{
R_9^B=\frac{1}{R_9^A}=\frac{\sqrt2}{r_9\sqrt{r_{10}}},\quad
g_9^B=g_9^A=\biggl(\frac{r_{10}}{2}\biggr)^{5/4}r_9^{-1/2},\quad
g_{10}^B=g_9^B\sqrt{R_9^B}=\frac{r_{10}}{2r_9}.}

Using these relations we are interested in the following 
three type IIB (0B) limits. 

\noindent a) $r_9\rightarrow 0$, $r_{10}\rightarrow 0$ with $R_9^B\to\infty$:
Here we obtain the type 0B string at finite or zero string
coupling $g_{10}^B=r_{10}/2r_9$.

\vskip0.2cm
\noindent b) $r_9\rightarrow 0$, $r_{10}\rightarrow\infty$ with 
$R_9^B\to\infty$:
This limit describes the strongly coupled type IIB string with 
$g_{10}^B\rightarrow\infty$.

\vskip0.2cm
\noindent c) $r_9\rightarrow\infty$, $r_{10}\rightarrow 0$ with
$R_9^B\to\infty$:
Now one is dealing with the weakly coupled type 0B string with 
$g_{10}^B\rightarrow 0$.

\vskip0.2cm

Let us now discuss how M-theory branes 
and the corresponding gauge theories can be incorporated into this picture.

\subsec{M2-branes}

The membrane solution (M2-brane) of 11-dimensional
supergravity plays a very important role
in the relation between string theory and M-theory.
Upon circle compactification from 11 to 10 dimensions  M2-branes with 
world volumes transversal to $x_{10}$ can be identified with the
IIA D2-branes. On the other hand, being wrapped around the eleventh dimension
the M2-brane provides the fundamental string of the IIA superstring.
Let us now discuss the fate of the M2-branes under the M-theory
compactification on $S^1\times S^1/(-1)^{F_s}S$, where the orbifold lies
in the $x_{10}$ direction. 
As a first set of membranes we introduce 2N M2-branes which are
completely unwrapped, so that
their worldvolumes are transversal to the compact two-dimensional orbifold.
Therefore this choice corresponds to the winding picture discussed before.
Of course we have to place N M2-branes at opposite positions on the circle. 
In order to obtain gauge theories the M2-branes are intersected by
another set of M2-branes which are wrapped around the orbi-circle in
$x_{10}$. Hence from the
string point of view these wrapped membranes correspond to the open strings
which are responsible for the gauge symmetry degrees of freedom.

In order to see what the various limits of M-theory parameters mean for
the corresponding gauge theories we need the  relations
between string theory and M-theory parameters, where we measure
all length scales in units of the 11-dimensional Planck length
$L_{11}$, which now appears explicitly in all formulas, i.e.
$r_9$ and $r_{10}$ are now dimensionful(!) quantities. Focussing
on type B quantities we obtain:
\eqn\twoba{
R_9^B=\frac{2L_{11}^3}{r_9r_{10}},\quad
g_{10}^B=
\frac{r_{10}}{2r_9}.}
Note that switching from $R_9^B$  measured  in string
units to the radius  measured in M-theory units
the following relation between the fundamental string scale $\alpha'$
and $L_{11}$ is required:
\eqn\tf{
T_F=\frac{1}{\alpha'}=\frac{r_{10}}{2L_{11}^3}.}
The ten-dimensional gravitational coupling constant is given by
\eqn\grav{ \kappa_{10}^2={L_{11}^9\over r_{10} } .}
In addition we also need the expression for the tension
of the solitonic D1-strings and D3-branes:
\eqn\td{
T_{D1}=\frac{T_F}{g_{10}^B}=\frac{r_9}{L_{11}^3},\quad\quad
T_{D3}=\frac{T_F^2}{g_{10}^B}=\frac{r_9\, r_{10}}{2L_{11}^6}.}

In the following we like to consider the gauge theories in the
limit $R_9^B\rightarrow\infty$, i.e. in the type B limit.
Moreover we like to study the behavior of the gauge theories in the
limit where the 11-dimensional Planck length is small, i.e.
$L_{11}=\rho\rightarrow 0$.

\vskip0.2cm
\noindent a) $r_9=\rho^a\rightarrow 0$ ($a>0$), $r_{10}=\rho^b\rightarrow 0$
($b>0$):

\vskip0.1cm\noindent
In order that $R_9^B\rightarrow\infty$ we have to require that $a+b>3$. Then 
this limit describes the non-supersymmetric type 0B string.
It contains N self-dual  D3-branes.
In order that the gauge theory modes completely decouple from the perturbative
as well as all non-perturbative stringy modes we demand in addition that
$T_F\rightarrow\infty$ and $T_{D1}\rightarrow\infty$.
This provides a further restriction on how $r_9$ and $r_{10}$ go to zero, 
namely
$a<3$ and also $b<3$.
Choosing the parameters in this way we have  a
non-supersymmetric
$U(N)\times U(N)$ gauge theory with 6 adjoint scalars and 4+4 bifundamental
fermions in four dimensions.
The gauge theory is weakly coupled, $g_{YM}=\sqrt{g_{10}^B}\rightarrow 0$,
if $b>a$. Moreover this non-supersymmetric gauge theory is expected to
possess an S-duality symmetry 
$g_{YM}\rightarrow 1/g_{YM}$, which is realized by the 
exchange of $r_9$ and $r_{10}$. Thus the strongly coupled 
non-supersymmetric $U(N)\times
U(N)$ gauge theory is obtained choosing $a>b$. Of course, one has to be very 
careful with such duality statements in the non-supersymmetric case,
as the lack of supersymmetry does not allow us to find more  supporting 
evidence for such a conjecture.

\vskip0.2cm
\noindent b) $r_9=\rho^a\rightarrow 0$ ($a>0$), 
$r_{10}=\rho^{-b}\rightarrow$ const. or $\infty$ ($b\geq 0$):

\vskip0.1cm\noindent
Choosing the two radii in this way we run towards supersymmetry restoration;
supersymmetry is completely restored for $r_{10}\rightarrow\infty$.
Again keeping $R_9^B$ large we need $a>3+b$, i.e. $a>3$.
Then in the supersymmetric limit the N M2-branes can be viewed as type IIB 
D3-branes with open strings ending on them. 
The open strings lead to 
${\cal N}=4$ supersymmetric $U(N)$ gauge theory at very strong
Yang-Mills gauge coupling $g_{YM}=\sqrt{g_{10}^B}\rightarrow\infty$.
However now the stringy modes
do not decouple anymore.
Although the fundamental string tension $T_F$ is still very large,
the D1 string tension $T_{D1}$ goes to zero, since $a>3$.
This means that we deal not only with strongly coupled
${\cal N}=4$ supersymmetric
$U(N)$ Yang-Mills, but also all D-stringy modes are present and do not 
decouple. We call this theory ${\cal N}=4$  MSYM.

\vskip0.2cm
\noindent c) $r_9=\rho^{-a}\rightarrow$ const. or $\infty$ ($a\geq 0$), 
$r_{10}=\rho^{b}\rightarrow 0$ ($b> 0$):

\vskip0.1cm\noindent
In this limit supersymmetry is completely broken.
Asking for $R_9^B\rightarrow\infty$ it implies  $b>3+a$, i.e. $b>3$.
Then  the N M2-branes can be viewed as self-dual type 0B 
D3-branes with non-supersymmetric 
 $U(N)\times U(N) $ gauge theory at very weak
Yang-Mill gauge coupling $g_{YM}=\sqrt{g_{10}^B}\rightarrow 0$.
Now the elementary stringy modes
do not decouple since $T_F\rightarrow 0$ whereas $T_{D1}\rightarrow\infty$. 
This means that we deal not only with classical 
non-supersymmetric
$U(N)\times U(N)$ Yang-Mills theory, 
but also all  elementary stringy modes are present and do not 
decouple. Let us call this theory ${\cal N}=0$ MYM theory. 
Under exchange of the two compact radii this case is
mapped to the limit described in b.), implying some strong-weak 
duality between these two supersymmetric resp. non-supersymmetric
 gauge theories. 
\vskip0.2cm

In all three type IIB (0B) limits the ten-dimensional gravitational
coupling vanishes and the tension of the D3-branes becomes infinite, so that
gravity and massive modes on the D3-branes decouple properly. 
If we take the M-theory discussion seriously, it tells us that 
the non-supersymmetric 
$U(N)\times U(N)$ gauge theory can be continued at very weak
and very strong coupling to two new branches of gauge   like theories,
where not all the stringy modes decouple from the dynamics. 
On one of these two branches the model can be deformed continuously 
to a supersymmetric model. 
Moreover, analogously to the bulk theory one might expect
a strong-weak duality for the non-supersymmetric
$U(N)\times U(N)$ gauge theory for finite
gauge coupling. In the large N limit this might really be true, as
the theory is conformal and the gauge coupling is a free parameter.

\subsec{M5-branes}

The solitonic solution dual to the membranes in 11-dimensional supergravity
are given by the M-theory 5-branes (M5-branes).
We will consider 2N M5-branes which are wrapped around both directions
of the compact space $S^1\otimes S^1/(-1)^{F_s}S$.
Hence from the string point of view we are in the momentum picture.
Therefore in nine dimensions the M5-branes correspond to 2N D3-branes.
These 2N M5-branes are intersected by M2-branes which correspond to
the open string sector in string theory.
In complete analogy to the unwrapped case, in the three type IIB (0B)
limits one obtains the following spectra.

\vskip0.2cm
\noindent a) $r_9=\rho^a\rightarrow 0$ ($a>0$), $r_{10}=\rho^b\rightarrow 0$
($b>0$):
We obtain N self-dual  D3 plus D3'-branes of
finitely coupled type 0B. Now we have non-supersymmetric
$U(N)\times U(N)$ gauge theory with 
the usual matter content.

\vskip0.2cm
\noindent b) $r_9=\rho^a\rightarrow 0$ ($a>0$), 
$r_{10}=\rho^{-b}\rightarrow$ const. or $\infty$ ($b\geq 0$):
Here we are dealing with the strongly coupled type IIB. After T-duality
in the $x_9$ direction the 2N M5-branes can be viewed as 2N D3-branes
with open strings ending on them. 
The corresponding gauge theory in the momentum picture 
is given by four-dimensional, 
${\cal N}=4$ supersymmetric $U(2N)$ gauge theory at very strong
Yang-Mill gauge coupling. However, not all D1-stringy modes decouple
and one does again  get ${\cal N}=4$  MSYM with gauge group $U(2N)$.

\vskip0.2cm
\noindent c) $r_9=\rho^{-a}\rightarrow$ const. or $\infty$ ($a\geq 0$), 
$r_{10}=\rho^{b}\rightarrow 0$ ($b> 0$):

Now one is dealing with N D3 plus N D3'-branes of weakly coupled 
type 0B string theory.
The corresponding gauge theory is weakly coupled ${\cal N}=0$ MYM with gauge 
group $U(N)\times U(N)$.

\vskip0.2cm

In the M5-brane scenario we get two new branches 
at very small and very large coupling, on which the supersymmetric
$U(2N)$ 
and the non-supersymmetric $U(N)\times U(N)$ gauge theory 
do still couple to some stringy modes.

\newsec{Conclusions}

In this paper we have shown that D-branes in freely acting orbifolds of
type II and  type 0 string constructions allow for a continuous
interpolation between ${\cal N}=4$ supersymmetric gauge theories
with $G=U(N)$ or $G=U(2N)$ gauge symmetry and a non-supersymmetric gauge
theory with gauge group $G=U(N)\times U(N)$, 6 adjoint scalars in each 
gauge group factor plus 4 Weyl fermions in the representations
$(N,\o{N})+(\o{N},N)$. The interpolation mechanism is of stringy nature
and can be realized 
varying the radii of the compact orbifold space. Therefore the coupling
of the open string degrees
of freedom to the modes of the closed string compactification are crucial
for the understanding of this mechanism.

We have also discussed how this scenario can be lifted to M-theory.
As a result of this discussion 
a non-supersymmetric gauge theory at weak coupling can be continuously
connected to a supersymmetric gauge theory like branch at strong coupling.
Another conclusion from 
this investigation was that the non-supersymmetric gauge theory itself
might have a strong-weak coupling duality. 
Let us emphasize, that duality conjectures for non-supersymmetric 
models are on less solid ground as compared to the supersymmetric case. 
Therefore, the M-theory results should be considered with some care.

It is clear that the discussion in this paper can be straightforwardly
generalized putting branes on transversal singularities or discussing
Hanany-Witten type of brane constructions.
Consider for example the case of D3-branes probing a transversal, non-compact
$\ZZ_K$ orbifold. 
Via T-duality this is equivalent
to $K$ NS 5-branes intersected by D4-branes \refs{\rkarch,\randreas}.
For the type II case, supersymmetry is broken
to ${\cal N}=2$ and the gauge group is now given by $G=U(N)^K$.
In the corresponding type 0 string the same construction leads
to a non-supersymmetric gauge theory with gauge group $\lbrack U(N)\times U(N)
\rbrack^K$ plus
certain matter fields. Again the large N $\beta$-function is the same as in 
the ${\cal N}=2$, type II parent model.
As before the freely acting orbifold construction 
implies that one can interpolate 
between these two ${\cal N}=2$ and ${\cal N}=0$ gauge theories.
For orbifolds or conifolds which break the amount of supersymmetry on the 
brane down to ${\cal N}=1$, an interpolating model can be build in an 
analogous way.

\bigskip\bigskip\centerline{{\bf Acknowledgements}}\pano
We thank C. Angelantonj, M. Green, A. Karch, I. Klebanov and
A. Sagnotti for useful discussions.
The  work is partially supported by the  European Commission TMR program
under the contract
ERBFMRXCT960090, in which the Humboldt-University at Berlin and the Ecole
Normale Superieur in Paris are associated.

\vfill\eject

\listrefs
\bye